\documentclass[twoside,a4paper,10pt]{article}
\usepackage{amsmath, amssymb, graphics}
\usepackage{amscd}
\usepackage{amssymb,amsmath,graphicx}

\begin{document}
\title{Monte Carlo sampling given a Characteristic Function: \\
\it Quantile Mechanics in Momentum Space}
\author{William T. Shaw\thanks{Corresponding author: Department of Mathematics
King's College, The Strand,
London WC2R 2LS, England; E-mail: william.shaw@kcl.ac.uk} and J. McCabe\thanks{Department of Mathematics
King's College London}}

\maketitle
\begin{abstract}
In mathematical finance and other applications of stochastic processes, it is frequently the case that the characteristic function may be known but explicit forms for density functions are not available. The simulation of any distribution is greatly facilitated by a knowledge of the quantile function, by which uniformly distributed samples may be converted to samples of the given distribution. This article analyzes the calculation of a quantile function direct from the characteristic function of a  probability distribution, without explicit knowledge of the density. We form a non-linear integro-differential equation that despite its complexity admits an iterative solution for the power series of the quantile about the median. We give some examples including tail models and show how to generate C-code for examples.\end{abstract}
\noindent
\bigskip
\noindent 

\pagestyle{myheadings}
\thispagestyle{plain}
\markboth{W.T. Shaw, J. McCabe}{Monte Carlo sampling given a characteristic function}

\section{Introduction}
The construction of Monte Carlo samples from a distribution is facilitated if one has a knowledge of the quantile function $w(u)$ of a distribution. If $F(x)$ is the cumulative distribution function associated with a continuous density, then the quantile $w(u)$ is the solution of the equation
\begin{equation}
F(w(u))=u\ . 
\end{equation}
A knowledge of the function $w(u)$ makes Monte Carlo simulation straightforward: given a random sample $U$ from the uniform distribution, a sample from the target distribution characterized by $f(x), F(x)$ is 
\begin{equation}
X = w(U)\ .
\end{equation}
This approach is especially useful when the $U$-variables arise as the components of a sampled copula, as then one {\it needs} the quantile functions of the marginals to create the marginal samples from the copula sample. But in general the method is useful, irrespective of whether copulae are involved. 

A full differential theory of quantiles based on a density function has been given recently by Steinbrecher and Shaw \cite{qmone}. The method is particularly easy to apply analytically when the density is of Pearson form and various power series solutions may be developed. This new paper addresses the case where the density is not known explicitly but where one just has its characteristic function. Such cases are of increasing interest, especially due to the role of jump diffusions and L\'{e}vy processes in mathematical finance. The methods developed here may also be applied to the cases of known density of more complicated form, as working via the characteristic function can turn out to be straightforward. We shall be able to give a detailed development here for the case where all integer moments of the characteristic function exist, i.e. where the probability density function is analytic about the origin.

\section{Quantile--characteristic function relationships}
If $f(x)$ is the probability density function for a real random variable $X$, the first order quantile ODE is obtained by differentiating Eqn. (1), to obtain:
\begin{equation}
f(w(u))\frac{dw(u)}{du} = 1, \label{fquant}
\end{equation}
where $w(u)$ is the quantile function considered as a function of $u$, with $0 \leq u \leq 1$. Applying the product rule with a further differentiation we obtain:
\begin{equation}
f(w(u))\frac{d^2w(u)}{du^2}+f'(w(u))\biggl(\frac{dw(u)}{du}\biggr)^2 = 0.
\end{equation}
This may be reorganized into the second order non-linear ODE given by Steinbrecher and Shaw \cite{qmone}:
\begin{equation}
\frac{d^2 w(u)}{du^2} = H(w(u)) \left(\frac{dw(u)}{du} \right)^2\ , \label{peaone}
\end{equation}
where
\begin{equation}
H(w) = -\frac{d\ }{dw} \log\{f(w)\}\ .
\end{equation}
and the simple rational form of $H(w)$ for many common distributions, particularly the Pearson family, allows analytical series solutions to be developed \cite{qmone}. Changes of variable may also be introduced, leading to a useful differential characterization of Cornish-Fisher expansions and candidates for quantile representations optimized for GPU computation \cite{qmtwo}.

Now suppose we do {\it not} have an explicit form for $f(w)$, but rather only have a characteristic function $\phi_X(t)$ defined by
\begin{equation}
\phi_X(t) = E[e^{itX}]
\end{equation}
So the theoretical density $f$ and $\phi_X$ are related as Fourier transform pairs with the conventions
\begin{equation}
\phi_X(t) = \int_{-\infty}^{\infty} f(x) e^{i t x} dx\ ,\ \ \ f(x) = \frac{1}{2\pi}\int_{-\infty}^{\infty} \phi_X(t) e^{-i t x} dt
\end{equation}
Combining these definitions allows us to write down the following three links between the quantile function and the characteristic function. First, the definition of the characteristic function can be written as 
\begin{equation}
\phi_X(t) = \int_{-\infty}^{\infty} e^{i t x} dF(x)
\end{equation}
where $F(x)$ is the cumulative distribution function associated with $f(x)$. We can then change variables to $u = F(x)$, with $x = w(u)$, to obtain a first quantile-characteristic link in the form:
\begin{equation}
\phi_X(t) = \int_0^1 e^{i t w(u)} du
\end{equation}
Second, by using the Fourier inversion theorem, we have, from Eqn.~(\ref{fquant}), 
\begin{equation}
\frac{dw(u)}{du}\int_{-\infty}^{\infty} \phi_X(t) e^{-i t w(u)} dt = 2\pi, 
\end{equation}
Third, and in the spirit of our previous approach, a further differentiation gives us
\begin{equation}
\frac{d^2w(u)}{du^2}\int_{-\infty}^{\infty} \phi_X(t) e^{-i t w(u)} dt  -i \biggl(\frac{dw(u)}{du}\biggr)^2\int_{-\infty}^{\infty} t \phi_X(t) e^{-i t w(u)} dt = 0. 
\end{equation}
These last two relationships may be combined to give a fourth identity in the form:
\begin{equation}
\frac{d^2w(u)}{du^2} =\biggl(\frac{dw(u)}{du}\biggr)^3   \frac{1}{2\pi} \int_{-\infty}^{\infty} i t \phi_X(t) e^{-i t w(u)} dt .
\end{equation}

This last equation will be the basis of our further analysis, though it is possible other efficient approaches may be developed from other such identities. We call it the characteristic quantile equation, or {\it CQE}. The CQE is a direct link between the characteristic and the quantile. If we can solve the CQE we can significantly short-cut what would be the normal three-stage route from the characteristic function to the quantile, and obtain Monte Carlo samples,  as illustrated in the diagram below.
$$
\CD
\phi_X(t) @>\text{Inverse Fourier transform}>> f(x) \\
@V\text{CQE solution?}VV @VV\text{Integration}V \\
w(u) @<<\text{Functional Inversion}< F(x)\\
@V\text{Sample uniform } UVV\\
X=w(U) \endCD
$$
It is quite often the case the the first step is analytically intractable, and if it does succeed the third step of functional inversion may be difficult. The diagonal route from the characteristic function direct to the CDF $F(x)$ {\it has} been considered, and we turn to this next.

\section{From the characteristic function to the distribution function}

The diagonal route from $\phi_X(t)$ to $F(x)$ is a well-trodden one. We refer to reader to Shephard (1991) \cite{shep} for a definitive survey on the results related to the ``Gil-Pelaez inversion'' formula \cite{gpi}.  We will discuss this result for two reasons. First we will need a special case for a result that anchors our quantile functions. Second, we wish to bring attention to the danger of using this result outside a carefully defined context - this comes to light if one considers an essentially {\it complex analytic} picture. 

The starting point is the elementary observation that the CDF $F(x)$ can be regarded as the convolution of the density $f(x)$ with the Heaviside step function $\theta(x)$ that is unity for $x>0$ and zero otherwise.
\begin{equation}
F(x) = \int_{-\infty}^x f(y) dy =  \int_{-\infty}^{\infty}\theta(x-y)f(y) dy 
\end{equation}
We therefore wish to exploit the convolution theorem, the Fourier transform of $F$ is given by
\begin{equation}
\hat{F}(t) = \phi_X(t) \hat{\theta}(t)
\end{equation}
Let us work out $\hat{\theta}(t)$ carefully:
\begin{equation}
 \hat{\theta}(t) =  \int_{-\infty}^{\infty}\theta(x)e^{i t x} dx  = \int_0^{\infty}\theta(x)e^{i t x} dx = \int_0^{\infty}e^{i t x} dx 
\end{equation}
The last integral is convergent if and only if $\Im(t)>0$, i.e. the transform exists and is holomorphic on the strict upper half-plane. In that case we obtain
\begin{equation}
 \hat{\theta}(t) =  \frac{1}{(-it)}
\end{equation}

In principle this means that an inversion integral involving $\hat{\theta}(t)$ should be carried out along a horizontal contour in the upper half plane $U = \{ z: \Im(z)>0\}$. Now consider $\hat{f}(t)$. This might not in fact be holomorphic (we will return to this later), but if it is it will normally be holomorphic on a horizontal strip $D = \{ z: B<\Im(z)<T\}$. {\it Provided} $D\cap  U$ is non-empty we pick a horizontal contour $C \in D\cap U$ and the inversion and convolution theorems give us
\begin{equation}
F(x) = \frac{i}{2 \pi} \ \int_C  \frac{\phi_X(t) e^{-i t x}}{t}
\end{equation}
Note that this whole idea requires that $D\cap U$ is non-empty. Furthermore, any attempt to move $C$ to the real axis might fail - this is {\it not} due to the presence of the factor $1/t$, which is in fact easily managed. Two cases of notable interest in the wider context of option-pricing are
\begin{itemize}
\item $f_1(x) = (e^x-1)^+$: the reader may verify by elementary integration that $D = \{ z: 1<\Im(z)\}$ and the real axis is well below $D\cap U= D$;
\item $f_2(x) = (1-e^x)^+$: the reader may verify by elementary integration that $D = \{ z: \Im(z)<0\}$, then $D\cap U= \emptyset$.
\end{itemize}
In these two examples $\hat{f_1} = \hat{f_2} = 1/(it-t^2)$ and the consideration of $D$ is all-important. However, the reader is right to note that these examples are somewhat pathological with regard to convergence - neither corresponds to the transform of a genuine PDF, whatever their wider relevance. 

Now suppose that $\phi_X(t)$ is holomorphic in a neighbourhood of the real axis. Then $D\cap U$ contains $C(\varepsilon)$, the horizontal contour $\{z: z = x + i \varepsilon; -\infty < x < \infty\}$ traversed from left to right. Then 
\begin{equation}
F(x) = \frac{i}{2 \pi}\int_{C(\varepsilon)}  \frac{\phi_X(t) e^{-i t x}}{t}
\end{equation}
We now deform $C(\varepsilon)$ to the contour  $C(\epsilon)$ (the mousehole contour) that is the union of the intervals 
\begin{equation}
\begin{split}
C_1 &= \{t: -\infty<t \leq \epsilon\}\\
C_2 &= \{t: t = \epsilon e^{i\theta}; \pi \geq \theta \geq 0 \}  \\
C_3 &= \{t:  \epsilon \leq t <  \infty\}\\
\end{split}
\end{equation}
and where $C_2$ is traversed clockwise.
Then we can write
\begin{equation}
F(x) = \frac{1}{2\pi}\int_{C_1} \frac{\phi_X(t)e^{-i t x}}{(-i t)}  dt+\frac{1}{2\pi}\int_{C_2} \frac{\phi_X(t)e^{-i t x}}{(-i t)} dt+\frac{1}{2\pi}\int_{C_3} \frac{\phi_X(t)e^{-i t x}}{(-i t)} dt
\end{equation}
and using the calculus of residues on $C_2$, we have
\begin{equation}
\lim_{\epsilon \rightarrow 0 }\frac{i}{2\pi}\int_{C_2} \frac{\phi_X(t)e^{-i t x}}{t} dt =  \frac{i}{2\pi}[-i\pi \phi_x(0)] = \frac{1}{2}\phi_X(0) = \frac{1}{2}
\end{equation}
We now change variables in $C_3$ with $t \rightarrow -t$ and  establish the result.
\begin{equation}
F(x) = \frac{1}{2} + \lim_{\epsilon \rightarrow 0}\frac{i}{2\pi}\int_{\epsilon}^{\infty}\frac{\phi_X(t)e^{-i t x}-\phi_X(-t)e^{i t x}}{t} dt
\end{equation}
Given that we have assumed $\phi$ is holomorphic we can expand the numerator and observe that the limit $\epsilon \rightarrow 0$ may be taken, finally giving us the Gil-Pelaez inversion formula:
\begin{equation}
F(x) = \frac{1}{2} + \frac{i}{2\pi}\int_{0}^{\infty}\frac{\phi_X(t)e^{-i t x}-\phi_X(-t)e^{i t x}}{t} dt
\end{equation}
Note that this result is the ``shadow'', evaluated along the real axis, of the result of equation (18) in a context where $C$ may be pushed arbitrarily close to the real axis. It is simply not true that this can be done with a general Fourier transform, as previously noted.

The category of Fourier transforms that we actually wish to deal with the set of transforms that arise from transforming probability density functions. This is a rather more subtle category, as the characteristic functions that arise are not necessarily in the category of functions holomorphic in a neighbourhood of the real axis. The contrast between the two-sided exponential distribution and the Cauchy distribution is useful. The two-sided exponential distribution has a transform with nice thick strip containing the real axis and bounded by two simple poles. The Cauchy distribution on the other hand has a {\it transform} of the form
\begin{equation}
\phi_X(t) = e^{-\lambda |t|}
\end{equation}
which is simply not holomorphic in a neighbourhood of the real axis (cf stable distributions in general). However, one can reconsider Eqn.~(24) in the altogether different setting of transforms of actually probability density functions, where the transforms are considered along the real axis only. We will not discuss this matter further here, and refer the reader instead to the work by Shephard \cite{shep}, which establishes that in this context the G-P inversion still works. However, the temptation to use Eqn.~(24) in any wider setting must be resisted, for the reasons already stated.

We now turn to the result that we shall need for the direct computation of quantile functions. We need this result to establish an origin for expansions. 

\medskip

\noindent{\bf Theorem 1 - the zero quantile location}
\medskip

\noindent The value of $u_0$ satisfying
\begin{equation}
w(u_0)=0
\end{equation}
is given by
\begin{equation}
u_0 = \frac{1}{2} + \frac{i}{2\pi}\int_{0}^{\infty}\frac{\phi_X(t)-\phi_X(-t)}{t} dt
\end{equation}
This resolved is now easily proved by substitution of $x=0$ in the inversion formula. We note that as expected $u_0 = 1/2$ when $f$, and hence $\phi_X$ is symmetric. 

\subsection{Examples of using the zero location theorem}
Consider an elementary case where we have not exploited the location symmetry on the simple Gaussian case to set the mean to zero. Then for a distribution with mean $\mu$ we have
\begin{equation}
\phi_X(t) = e^{i\mu t - t^2/2}
\end{equation}
The location theorem gives us, following some first simplification:
\begin{equation}
u_0 = \frac{1}{2} - \frac{1}{2\pi} \int_0^{\infty} e^{-t^2/2} \frac{2\sin(\mu t)}{t} dt  = \frac{1}{2} - \frac{1}{2} {\rm erf}\biggl(\frac{\mu}{\sqrt{2}}\biggr) = N(-\mu)
\end{equation}
where $N(x)$ is the normal CDF. This of course is the correct result.  
\subsection{Stable distribution}
Next, consider a stable distribution with
\begin{equation}
\phi_X(t) = \exp\{-|t|^{\alpha} (1-i\beta{\rm sign(t)}\Phi)  \}
\end{equation}
and $\Phi = \tan(\alpha \pi/2)$. The location theorem gives us
\begin{equation}
u_0 =  \frac{1}{2} - \frac{1}{\pi} \int_0^{\infty} e^{-t^{\alpha}} \frac{\sin(\beta \Phi t^{\alpha})}{t} 
\end{equation}
The integral evaluates to give us
\begin{equation}
u_0 =  \frac{1}{2} - \frac{1}{\pi\alpha} \arctan(\beta \Phi) \ .
\end{equation}
Note than when $\beta = 0$ we obtain $u_0 = 1/2$ and that when $\beta = 1$ we obtain $u_0 = 0$. In the latter case the distribution only exists for non-negative $X$. When $\alpha<1$ zero is not an interior point of the support of $X$. Our location formula behaves in the right way and we can also see that as $\beta \rightarrow -1$ then $u_0 \rightarrow 1$. 

The point of these observations is that we can get a base point for a series expansion of the quantile by the computation of an integral. In the stable case we can see that the result is a simple ``closed-form'' formula.

\section{Formal solution of the characteristic quantile equation}
We have a non-linear integro-differential equation linking the quantile to the characteristic function. Inspection of the CQE indicates that we can differentiate it , then eliminate $w''(u)$ from the right hand side by using Eqn. (13) again. Let's do this once:
\begin{equation}
w'''(u) = [w'(u)]^3  \frac{1}{2\pi} \int_{-\infty}^{\infty} i t (-i t) \phi_X(t) e^{-i t w(u)} dt + 3[w'(u)]^2 w''(u)  \frac{1}{2\pi} \int_{-\infty}^{\infty} i t \phi_X(t) e^{-i t w(u)} dt .
\end{equation}
Use of the CQE and some simplification leads to 
\begin{equation}
w'''(u) = [w'(u)]^4  \biggl[\frac{1}{2\pi} \int_{-\infty}^{\infty} t^2 \phi_X(t) e^{-i t w(u)} dt  + 3 w'(u) \frac{1}{4 \pi^2} \biggl(\int_{-\infty}^{\infty} i t \phi_X(t) e^{-i t w(u)} dt\biggr)^2  \biggr]
\end{equation}

Some experimentation lead us to write
\begin{equation}
w^{(n)}(u) = [w'(u)]^{n+1} P_n[w'(u), w(u)]
\end{equation}
where $P_n[w'(u), w(u)]$ is a polynomial in $w'(u)$  of degree $n-2$ whose coefficients involve integrals of $w(u)$. The operation of differentiation and resubstitution gives the recurrence identity:

\begin{equation}
P_{n+1}[x, w] = (n+1) x P_2[w] P_n[x,w] + x^2 P_2[w] \frac{\partial\ }{\partial x}P_n[x,w] + \frac{\partial \ }{\partial w}P_n[x,w] 
\end{equation} 
where
\begin{equation}
P_2[x,w] = P_2[w]=\frac{1}{2\pi} \int_{-\infty}^{\infty} i t \phi_X(t) e^{-i t w} dt 
\end{equation}
We can do an inductive verification of this relation and note its correctness when $n=2$.

\subsection{The analytic assumption}
In carrying out the repeated differentiation of the CQE we are assuming that all derivatives of $w(u)$ exist and that we may form a convergent power series with which to represent the solution. The operation of repeated differentiation on the quantile will call up integrals of the form
\begin{equation}
M_k = \frac{1}{2\pi}\int_{-\infty}^{\infty} t^k \psi(t) dt
\end{equation}
and these are related to the derivatives of the density function (if they exist) by
\begin{equation}
f^{(k)}(0) = (-i)^k M_k
\end{equation}
from the Fourier inversion theorem. What our series solution is using the characteristic function to invert a Taylor series for the density about the origin without explicitly calculating it. In our initial solution to the problem, this series must exist in some form - other methods must be used if no such series exists.

\subsection{Symmetric distributions on the entire real line}

How might we use this to solve our problem? Let's consider an interesting special but rich case, where the density is symmetric about the origin, and the random variables $X$ may take all real values. In the symmetric case $w(1/2)=0$ and we can consider a power series solution about the median. Inspection of the integrals obtained so far gives us $w''(1/2)=0$ and
\begin{equation}
w'''(1/2) =  [w'(1/2)]^4  \frac{1}{2\pi} \int_{-\infty}^{\infty} t^2 \phi_X(t) dt 
\end{equation}
The normalization $w'(1/2)$ is easily determined by the condition:
\begin{equation}
\frac{1}{w'(1/2)} =  \frac{1}{2\pi} \int_{-\infty}^{\infty} \phi_X(t) dt 
\end{equation}
One observes that the formal solution for the quantile as a power series around the median is in principle completely determined by a knowledge of the even ``moments'' of the characteristic function , as these supply the coefficients of that power series. Furthermore they do so directly, without having to establish the density $f$ or CDF $F$.  In this symmetric case we can write down the complete solution for the quantile as the series
\begin{equation}
w(u) = v+ w'(1/2)\sum_{k=1}^{\infty}P_{2k+1}[w'(1/2), 0] \frac{v^{2k+1}}{(2k+1)!}
\end{equation}
where $v = w'(1/2)(u-1/2)$.
The main issue that remains is the computation of the $P$-coefficients in terms of ``moments'' of the characteristic function, by the solution of the recurrence identity Eqn. (17) with the initial condition of Eqn. (18). One also expects such power series to become increasingly awkward as one moves into the tails, so we expect to have to supplement the model with a special treatment of the tail.

\subsection{Locational invariance of the CQE}
The CQE has a notable, if rather obvious symmetry. It is invariant under the transformations
\begin{equation}
\phi_X(t) \rightarrow e^{i t a}\phi_X(t)\ , w(u) \rightarrow w(u)+a \ .
\end{equation}
This of course corresponds to a shift in the density
\begin{equation}
f(x) \rightarrow f(x-a)\ ,
\end{equation}
as one would expect.

\subsection{Scale invariance of the CQE}
The transformation
\begin{equation}
t \rightarrow c t\ ,\ \ \phi_X(t) \rightarrow \phi_X(c t)\ , \ \ w(u) \rightarrow w(u)/c\ .
\end{equation}
also generates a symmetry of the CQE. 

The scale and location invariance may often be used to transform the problem to one in standard form with fewer parameters. 

\subsection{Asymmetric distributions on $(-\infty, \infty)$}
Once any relevant locational and scale transformations have been exploited, a distribution may remain asymmetric. In this case the choice of an origin for expansion must be resolved. In the symmetric case $w(1/2)=0$ is the median so we might consider whether the try to still expand about the median or about the point $u_0$ chosen so that $w(u_0)=0$. The advantage of the latter approach is that we can do the expansion again in terms of simple ``moments'' of the characteristic function, and we can compute an expression for $u_0$. We have the zero quantile location as a special case of the Gil-Pelaez inversion formula
\begin{equation}
u_0 = \frac{1}{2} + \frac{i}{2\pi}\int_{0}^{\infty}\frac{\phi_X(t)-\phi_X(-t)}{t} dt
\end{equation}

\subsection{Formal solution for the asymmetric case}
Having found $u_0$ using the methods just described, we then have, as before,
\begin{equation}
\frac{1}{w'(u_0)} =  \frac{1}{2\pi} \int_{-\infty}^{\infty} \phi_X(t) dt 
\end{equation}\begin{equation}
w(u) = v+ w'(u_0)\sum_{k=2}^{\infty}P_k[w'(u_0), 0] \frac{v^{k}}{k!}
\end{equation}
where $v = w'(u_0)(u-u_0)$.

\subsection{Exponential asymmetry}
For a significant class of distributions asymmetry in the system is introduced by an exponential scaling. That is, for a real parameter $\beta$, 
\begin{equation}
f(x;\beta) = \frac{c(\beta)}{c(0)} f_S(x) e^{\beta x}
\end{equation}
where $c()$ is a normalization function and $f_S$ is a symmetric case. It is evident by the shift theorem that
\begin{equation}
\psi(t;\beta) = \frac{c(\beta)}{c(0)} \psi_S(t-i\beta)
\end{equation}
and hence the quantile must be determined only by integrals of the symmetric characteristic function $\psi_S$. We can make this explicit by a complex shift of the integration contour\footnote{We can do this by Cauchy's theorem, given that $\beta$ must be constrained so that $f$ remains $L^1$ integrable and the transform does not develop singularities} in the CQE with the change of variables $t = p+i\beta$. We obtain the modified CQE linking the full quantile with the symmetric $\psi_S$:
\begin{equation}
\frac{d^2w(u)}{du^2} =\biggl(\frac{dw(u)}{du}\biggr)^3   \frac{c(\beta)}{2\pi c(0)} e^{\beta w(u)}\int_{-\infty}^{\infty} (ip-\beta) \psi_S(t) e^{-ip w(u)} dp .
\end{equation}

\section{Series solution of the CQE - the symmetric case}
Let us summarize the problem in the symmetric case. Given a characteristic function $\phi_X(t)$, we need to solve the iteration scheme:
\begin{equation}
P_{n+1}[x, w] = (n+1) x P_2[w] P_n[x,w] + x^2 P_2[w] \frac{\partial\ }{\partial x}P_n[x,w] + \frac{\partial \ }{\partial w}P_n[x,w] 
\end{equation}
with the initial condition: 
\begin{equation}
P_2[x,w] = P_2[w]=\frac{1}{2\pi} \int_{-\infty}^{\infty} i t \phi_X(t) e^{-i t w} dt 
\end{equation}
The power series of the quantile function about the median $w(1/2)=0$  is then
\begin{equation}
w(u) = v+ w'(1/2)\sum_{k=1}^{\infty}P_{2k+1}[w'(1/2), 0] \frac{v^{2k+1}}{(2k+1)!}
\end{equation}
where $v = w'(1/2)(u-1/2)$ and
\begin{equation}
\frac{1}{w'(1/2)} =  \frac{1}{2\pi} \int_{-\infty}^{\infty} \phi_X(t) dt \ .
\end{equation}
The solution of the iteration is a rather mindless process that is best automated in a symbolic computation environment, and we employed Mathematica to carry out the analysis. We summarize the results. Let
\begin{equation}
E_k = \frac{1}{2 \pi}\int_{-\infty}^{\infty} t^{2k} \phi_X(t) dt
\end{equation}
denote the $k$'th normalized even moment. We have already observed that
\begin{equation}
P_3[x,0] = E_1
\end{equation}
Symbolic iteration then yields further terms as follows, where we abbreviate $P_k[x,0] = p_k$,
\begin{equation}
\begin{split}
p_5 &= 10 x E_1^2 - E_2 \\
p_7 &= 280 x^2 E_1^3-56 x E_2 E_1+E_3\\
p_9 &=15400 x^3 E_1^4-4620 x^2 E_2 E_1^2+x \left(126 E_2^2+120 E_1 E_3\right)-E_4\\
p_{11} &=1401400 x^4 E_1^5-560560 x^3 E_2 E_1^3+x^2 \left(17160 E_3 E_1^2+36036 E_2^2 E_1\right)-x \left(792 E_2 E_3+220 E_1 E_4\right)+E_5\\
p_{13}&=190590400 x^5 E_1^6-95295200 x^4 E_2 E_1^4+x^3 \left(3203200 E_3 E_1^3+10090080 E_2^2
   E_1^2\right)\\
   &\ \ \ -x^2 \left(126126 E_2^3+360360 E_1 E_3 E_2+50050 E_1^2
   E_4\right)+x \left(1716 E_3^2+2002 E_2 E_4+364 E_1 E_5\right)-E_6\\
p_{15}&=  36212176000 x^6 E_1^7-21727305600 x^5 E_2 E_1^5+x^4 \left(775975200 E_3 E_1^4+3259095840
   E_2^2 E_1^3\right)\\
   &\ \ \    +x^3 \left(-13613600 E_4 E_1^3-147026880 E_2 E_3
   E_1^2-102918816 E_2^3 E_1\right)\\
      &\ \ \ +x^2 \left(123760 E_5 E_1^2+1166880 E_3^2
   E_1+1361360 E_2 E_4 E_1+2450448 E_2^2 E_3\right)\\
      &\ \ \ +x \left(-11440 E_3
   E_4-4368 E_2 E_5-560 E_1 E_6\right)+E_7
\end{split}\label{firstfew}
\end{equation}
The display here of further terms would become rather unwieldy. In any case, such expressions are best stored symbolically for subsequent simplification for a particular distribution. We do note that this is a {\it one-off} computation that once done to high order in a symbolic computation environment can then be transferred to another computer environment for implementation along with computation of the characteristic moments.

\subsection{Testing the approach with the normal distribution}
This is one where a complete algebraic characterization of the median power series has been given \cite{qmone}.
\begin{equation}
\phi_X(t) = e^{-t^2/2}
\end{equation}
All the relevant integrals exist:
\begin{equation}
E_{k} = \frac{1}{2\pi}\int_{-\infty}^{\infty} t^{2n}  e^{-t^2/2} dt =\frac{1}{\pi}2^{n-\frac{1}{2}} \Gamma \left(n+\frac{1}{2}\right)
\end{equation}
and in particular
\begin{equation}
w'(1/2) = \sqrt{2\pi}
\end{equation}
Then the use of the series above, Eqns. (27)-(31), and some simplification yields the normal quantile in the form:
\begin{equation}
\begin{split}
\sqrt{2 \pi } \left(u-\frac{1}{2}\right)+\frac{1}{3} \sqrt{2} \pi ^{3/2}
   \left(u-\frac{1}{2}\right)^3+\frac{7 \pi ^{5/2}
   \left(u-\frac{1}{2}\right)^5}{15 \sqrt{2}}+\frac{127 \pi ^{7/2}
   \left(u-\frac{1}{2}\right)^7}{315 \sqrt{2}}+\frac{4369 \pi ^{9/2}
   \left(u-\frac{1}{2}\right)^9}{11340 \sqrt{2}}\\+\frac{34807 \pi ^{11/2}
   \left(u-\frac{1}{2}\right)^{11}}{89100
   \sqrt{2}}+O\left(\left(u-\frac{1}{2}\right)^{13}\right)
   \end{split}
   \end{equation}
Many more terms can be computed using some computer algebra. The generation of this particular series is however best handled with the methods of \cite{qmone}, where a purely algebraic recursion may be used instead of Eqn. (27). 
\subsection{Testing on the Student distribution}
For a Student $t$ distribution with $n$ degrees of freedom the characteristic function is given by
\begin{equation}
\phi_{T_n}(t) = \frac{2^{1-\frac{n}{2}} n^{\frac{n+2}{4}-\frac{1}{2}} |t|^{n/2}
   K_{\frac{n}{2}}\left(\sqrt{n} |t|\right)}{\Gamma
   \left(\frac{n}{2}\right)}
\end{equation}
See e.g. the detailed discussion by Hurst \cite{hurst} for an elegant derivation of this. The normalization is
\begin{equation}
x = w'(1/2) = \frac{ \sqrt{n\pi } \Gamma \left(\frac{n}{2}\right)}{\Gamma
   \left(\frac{n+1}{2}\right)}\ .
\end{equation}
All the relevant characteristic moments exist in the form
\begin{equation}
E_k = \frac{4^k n^{-k-\frac{1}{2}} \Gamma
   \left(k+\frac{1}{2}\right) \Gamma
   \left(k+\frac{n}{2}+\frac{1}{2}\right)}{\pi  \Gamma
   \left(\frac{n}{2}\right)}
\end{equation}
The method leads to the Student quantile series as given in \cite{shawjcf06} and as extended in \cite{qmone}. 
\section{Symmetric Distributions of interest}
We may consider distributions whose density is known, or not known, in explicit form. \subsection{Symmetric stable distribution}
These have been extensively discussed - see Nolan. Employing location and scale invariance allows us to focus attention on the symmetric case in standard form:
\begin{equation}
\phi_X(t) = e^{-|t|^{\alpha}}
\end{equation}
The tail behaviour is given to leading order by
\begin{equation}
f(x) \sim \frac{\alpha \sin(\pi\alpha/2)\Gamma(\alpha)}{\pi|x|^{(1+\alpha)}}
\end{equation}
All the relevant integrals exist:
\begin{equation}
E_k = \int_{-\infty}^{\infty} t^{2k} e^{-|t|^{\alpha}} dt = \frac{2}{\alpha} \Gamma\biggl(\frac{2k+1}{\alpha} \biggr)
\end{equation}
and in particular
\begin{equation}
w'(1/2) = \frac{\pi}{\Gamma[1+1/\alpha]}
\end{equation}

\subsection{Symmetric Generalized Hyperbolic}
This is a case, SGH, where the density is known in closed form. Nevertheless, while repeated differentiation of the density is possible for certain parameters, it turns out to be much more tractable to work in momentum space. The SGH characteristic function is given by
\begin{equation}
\phi_X(t) = \biggl(\frac{\alpha^2}{\alpha^2+z^2} \biggr)^{\lambda/2}\frac{K_{\lambda}(\delta\sqrt{\alpha^2+z^2})}{K_{\lambda}(\alpha \delta)}
\end{equation}
and is associated with the density function
\begin{equation}
f(x) = \frac{(\alpha/\delta)^{\lambda}}{\sqrt{2\pi}K_{\lambda}(\alpha \delta)}\frac{K_{\lambda-1/2}(\alpha\sqrt{\delta^2+x^2})}{(\sqrt{\delta^2+x^2}/\alpha)^{1/2-\lambda}}
\end{equation}
We assume that $\alpha\geq0$ and $\delta \geq0$. 
There are several special cases:
\begin{itemize}
\item Student: $\lambda = -\nu/2$, $\delta = \sqrt{\nu}$, $\alpha \rightarrow 0_+$.
\item Ordinary hyperbolic: $\lambda = 1$;
\item NIG: $\lambda = -1/2$;
\item Symmetric Variance Gamma: $\delta \rightarrow 0_+$.
\end{itemize}
The value of the density at the origin is given by
\begin{equation}
f(0) = \frac{(\alpha/\delta)^{\lambda}}{\sqrt{2\pi}K_{\lambda}(\alpha \delta)}\frac{K_{\lambda-1/2}(\alpha\delta)}{(\delta/\alpha)^{1/2-\lambda}} = \sqrt{\frac{\alpha}{2 \pi \delta}}\frac{K_{\lambda-1/2}(\alpha\delta)}{K_{\lambda}(\alpha\delta)}
\end{equation}
and so
\begin{equation}
w'(0) =\sqrt{\frac{2 \pi \delta}{\alpha}}\frac{K_{\lambda}(\alpha\delta)}
{K_{\lambda-1/2}(\alpha\delta)}\end{equation}
The characteristic moments are given by (exploiting the symmetry)
\begin{equation}
E_k = \frac{1}{\pi} \int_0^{\infty}z^{2k}\biggl(\frac{\alpha^2}{\alpha^2+z^2} \biggr)^{\lambda/2}\frac{K_{\lambda}(\delta\sqrt{\alpha^2+z^2})}{K_{\lambda}(\alpha \delta)}
\end{equation}
Making the change of variable $p = \sqrt{\alpha^2+z^2}$
\begin{equation}
E_k = \frac{1}{\pi} \int_{\alpha}^{\infty}(p^2-\alpha^2)^{(k-1/2)}\alpha^{\lambda}p^{1-\lambda}\frac{K_{\lambda}(\delta p)}{K_{\lambda}(\alpha \delta)}
\end{equation}
This integral can be evaluated by Mathematica \cite{math}, which returns the value
\begin{equation}
E_k = 2^{k-1/2}\Gamma(k+1/2)\biggl(\frac{\alpha}{\delta}  \biggr)^{k+1/2} \frac{K_{1/2+k-\lambda}(\alpha\delta)}{\pi K_{\lambda}(\alpha\delta)}
\end{equation}
subject to computer-generated constraints $k>0, \alpha>0, \delta>0, \lambda<2$.

\subsection{A Harder Example: the Levy Stochastic Area}
The distribution of the Levy Stochastic Area (LSA) is of considerable interest from the point of view of stochastic analysis and high order Monte Carlo simulation. We base our analysis on the approach of Schmitz \cite{klaus}. For a pair of Brownian motions spanning a time $\Delta t$ the LSA is given by
\begin{equation}
L(\Delta t) = \int_0^{\Delta t}(W_1(t) dW_2(t) - W_2(t) dW_1(t))
\end{equation}
Its characteristic function, conditional on a known value of  $R^2 = W_1^2(\Delta t) +   W_2^2(\Delta t)$, is known and given by
\begin{equation}
\begin{split}
\phi_L(z) &= \phi_X(z) \phi_Y(z)\\
\phi_X(z) &= \frac{z \Delta t}{\sinh(z \Delta t)}\\
\phi_Y(z) &= \exp\biggl[ -\frac{R^2}{2\Delta t}\big[z \Delta t \coth(z \Delta t)-1\bigr]   \biggr]
\end{split}
\end{equation}
where $W_i(0)=0$. There are two components. When the path goes nowhere (loop) $R=0$ and this is given by the random variable $X$. Then there is a further contribution from $Y$. This split is convenient as $X$ has a trivial quantile function:
\begin{equation}
Q_X(u) = \frac{\Delta t}{\pi} \log\biggl( \frac{u}{1-u}\biggr)
\end{equation}
Readers should consult \cite{klaus} for details and an extensive discussion of this entity and further references. The difficult part is finding $Q_Y(u)$. The characteristic function is clearly symmetric. The properties of the system are best understood by first extracting the time-scaling. From the definition $L$ and $X$ are both $O(\Delta t)$ times some time-scale-independent random variable.  So we set $Y = \Delta t P$ and $R^2 = r^2\Delta t$, so that $r$ is the distance gone by a Brownian motion in {\it unit time}.  Making the change of variables shows that
\begin{equation}
f_Y(x) = \frac{1}{\Delta t} g\biggl(\frac{x}{\Delta t} \biggr)
\end{equation}
and the characteristic function of $P$ is just
\begin{equation}
\phi_P(s) = \exp[-\frac{r^2}{2}(s \coth s - 1)]
\end{equation}
and the associated density function is
\begin{equation}
g_P(p) = \frac{1}{2\pi}\int_{-\infty}^{\infty}\exp[i s p-\frac{r^2}{2}(s \coth s - 1)]ds
\end{equation}
When $s$ is small or $p$ is large the hyperbolic function may be expanded to give
\begin{equation}
\phi_P(s) \sim \exp\biggl(-\frac{r^2 s^2}{6} \biggr)
\end{equation}
so that samples of $P$ may be approximately constructed with
\begin{equation}
P \sim r \sqrt{\frac{1}{3}}Z
\end{equation}
where $Z$ is a Gaussian random variable obtained by a quantile or other means. 

\section{Implementation}
There are a number of issues to address when it comes to implementation. These include
\begin{itemize}
\item The management of a symbolic differential recursion;
\item The choice of a computer language in which to solve the problem;
\item Tail management;
\item Error analysis;
\item Implementation in legacy or specialized computer environments.
\end{itemize}

The implementation of the differential recursion summarized in Section 4 appears to necessitate a modern symbolic computation environment and we have implemented it in a few lines of {\it Mathematica} \cite{math}. But it is important to point out that this needs only to be done {\it once} and we have solved it for the symmetric case, the first few terms being given explicitly in Eqn.~(\ref{firstfew}) up to $k=60$? The computation becomes labourious for $k>50$. The solution can be saved and then exploited. We can carry out the remain parts of the calculation entirely in {\it Mathematica}, as it can evaluate and store the characteristic moments and generate the relevant power series for the central portion of the quantile. 

However, while this constitutes a complete solution for a truncated central power series, many potential users may need the ability to migrate the model to another computer language. For example, financial specialists may require C/C++ implementations. Statisticians might want an implementation in {\it R}. Some scientific and engineering applications will require a FORTRAN$XX$, where probably $XX\geq90$. We can consider trying to migrate the solution to another language at various stages. In this first implementation we will consider a {\it late migration} model, with C/C++ as a target. This offers the compromise of getting {\it Mathematica} to do all the hard symbolic work, but we can output very simple C code to embed in a function to do the work in another program, once the parameters have been fixed. Migration one stage earlier would require calling a number of special functions in C, in particular various types of Bessel function. There are libraries to do this. Migration a further step back would require the implementation in C of the solution for the $p_k$. In principle this could be carried out with the {\tt CForm} command used below in the late migration approach.  We have not considered how the symbolic recursion itself might be solved directly in other languages. 

The error analysis is difficult as we do not know the benchmark, apart from special cases. We can give estimates based on this. Furthermore, if some form of expression for the cumulative distribution function (CDF), $F$,  is available, we can certainly estimate the round-trip error
\begin{equation}
RTE(u) = F[(Q(u)]-u
\end{equation}
and from this, given a density $f$, we can estimate the quantile (EQE) error as
\begin{equation}
EQE(u) = (F[Q(u)]-u)/f[Q(u)]
\end{equation}
which is of course the Newton-Raphson correction. 

In each case the central series will almost almost require a tail model. The power series will work well in a region $\epsilon<u<1-\epsilon$ but the region $1-\epsilon \geq u \geq 1$ and its mirror will need separate management.

\subsection{Solution of the differential recursion}
This takes place as follows. First we define $P_2$:
\begin{verbatim}
P[2, x_, w_] := 1/(2 Pi) Integrate[I t Phi[t] Exp[-I t w], {t, -Infinity, Infinity}]
\end{verbatim}
Subsequent terms are then, for the general case without symmetry, given by
\begin{verbatim}
P[n_, x_, w_] := 
P[n, x, w] = Expand[n*x*P[2, x, w]*P[n - 1, x, w] + 
                     x^2 D[P[n - 1, x, w], x] P[2, x, w] + D[P[n - 1, x, w], w]  ]\end{verbatim}
This is the key calculation. Two other operations rewrite the integrals that result in terms of the $E_k$ and apply symmetries. The program that implements all this is given in Appendix A, where we exploit the symmetry to give a program to just work out every other term. The results are stored in a file and we have computed as far as $p_{71}$ in symbolic form. 
 
 \subsection{Example 1: Symmetric Stable, central series for quantile}

The symbolic computation of the $p_k$ may be combined with the evaluation of the $E_k$ and $w'(0)$ given in Section 6.1, to give the central power series for the quantile. The first few terms are given by:
\begin{equation}
w(u)=\frac{\pi  \alpha  \left(u-\frac{1}{2}\right)}{\Gamma
   \left(\frac{1}{\alpha }\right)}+\frac{\pi ^3 \alpha ^3 \Gamma
   \left(\frac{3}{\alpha }\right) \left(u-\frac{1}{2}\right)^3}{6
   \Gamma \left(\frac{1}{\alpha }\right)^4}-\frac{\left(\pi ^5 \alpha
   ^5 \left(\Gamma \left(\frac{1}{\alpha }\right) \Gamma
   \left(\frac{5}{\alpha }\right)-10 \Gamma \left(\frac{3}{\alpha
   }\right)^2\right)\right) \left(u-\frac{1}{2}\right)^5}{120 \Gamma
   \left(\frac{1}{\alpha
   }\right)^7}+O\left(\left(u-\frac{1}{2}\right)^7\right)
\end{equation}
This expression is useful for some basic checking in the central zone but is not accurate enough for serious Monte Carlo simulation. One interesting check is to ask the computer to symbolically invert the series, and then differentiate the result. We obtain
\begin{equation}
f(x) = \frac{\Gamma \left(\frac{1}{\alpha }\right)}{\pi  \alpha
   }-\frac{\Gamma \left(\frac{3}{\alpha }\right) x^2}{2
   (\pi  \alpha )}+\frac{\Gamma \left(\frac{5}{\alpha
   }\right) x^4}{24 \pi  \alpha }+O\left(x^6\right)
\end{equation}
Readers may recognize this as Bergstrom's central series \cite{bergstrom} for the density function, which is a nice point of verification\footnote{The use of {\it Mathematica}'s {\tt InverseSeries} command to compute the quantile from a density series may also be invoked, but is not practical for very large series.} For high precision work, we need more terms. The stored internal representation in Mathematica may be used, and in particular to test the precision of the result for the known Gaussian and Cauchy cases. In the Gaussian case the {\it standard} normal quantile is $1/\sqrt{2}$ times the stable quantile with $\alpha=2$. We may compare the results with {\it Mathematica}'s internal high-precision routines. We will plot the precision in the form 
\begin{equation}
\log_{10}\biggl[\frac{{\text Power\ Series\ from\ Stable\ Distribution:} \alpha=2}{\sqrt{2}\times({\text Internal\ High-Precision\ Quantile)}}-1 \biggr]
\end{equation}
\begin{figure}[hbt]
\centering
\includegraphics[scale=0.8]{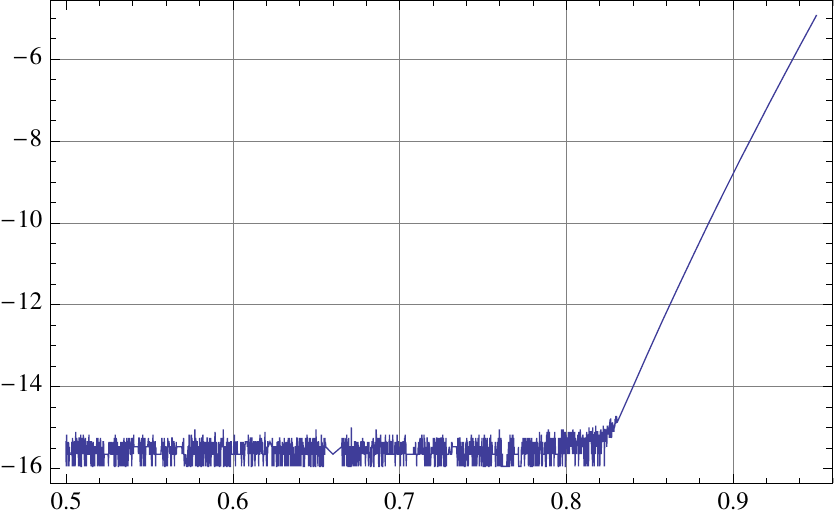}
\caption{Precision of the Gaussian stable quantile: $0.5\leq u\leq0.95$}
\end{figure}
\begin{figure}[hbt]
\centering
\includegraphics[scale=0.8]{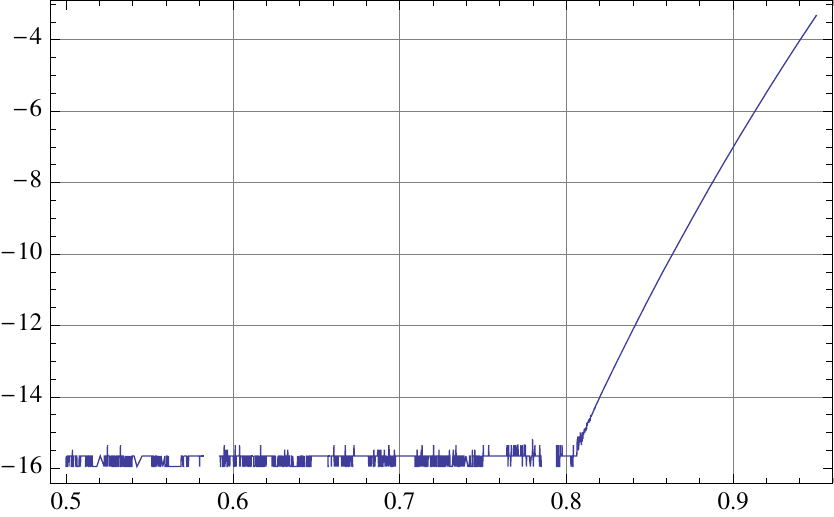}
\caption{Precision of the Cauchy stable quantile: $0.5\leq u\leq0.95$}
\end{figure}
\begin{figure}[hbt]
\centering
\includegraphics[scale=0.8]{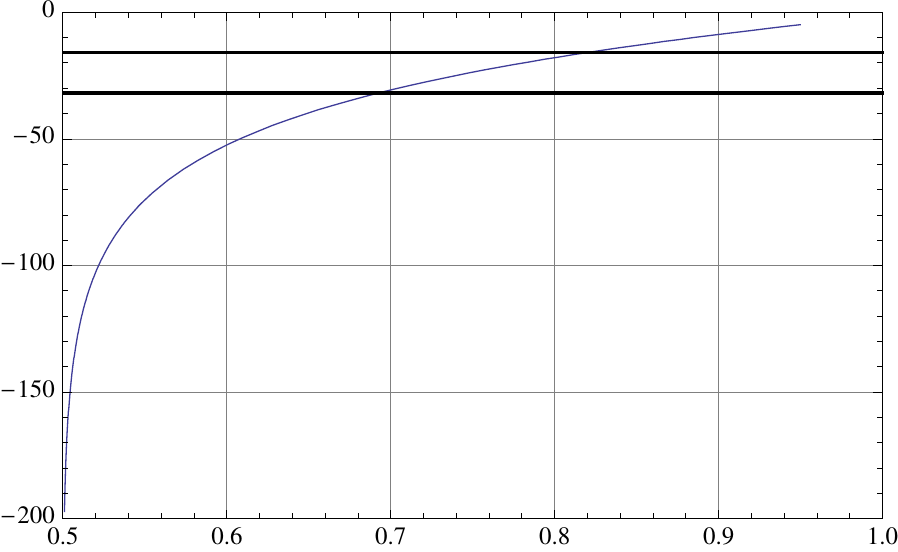}
\caption{Full precision of the Gaussian stable quantile: $0.5\leq u\leq0.95$}
\end{figure}

We note that the error is about machine precision level for $0.5\leq u < 0.84$ and then rises into the tail, remaining at an acceptable level until some point around $0.94$. Similar results apply to the Cauchy case, where no factor of $1/\sqrt{2}$ is needed, and the error grows slightly faster than in the Gaussian case. 
Note also that the machine precision oscillations in the plot are due to restricting attention to compiled results within the plotting routine - the series becomes much more accurate than one part in $10^{16}$ as one approaches the median. We can illustrate this by asking for one of the comparisons to be re-done in arbitrary precision mode. The result is shown in Figure 3, with quad- and double-precision limits shown as horizontal lines at $-32$ and $-16$.

\subsection{The stable quantile with $\alpha=3/2$: migration to C/C++}
Having tested the symbolic stable quantile on known cases $\alpha=2,1$, we now consider the intermediate case $\alpha = 3/2$. In this case we may take our symbolic representation and convert it into working C/C++ code. There are some helpful intermediate steps. First, we introduce variables $v=2u-1$ and $w=v^2$. Second, we use the symbolic computation engine to write the system in the standard nested multiplication form for polynomials (truncated power series). In {\it Mathematica} this involves the use of the {\tt HornerForm} function. Finally we use {\tt CForm} to output the code. The output of this takes the form:
\begin{verbatim}
v*(1.74002161967547716294123 + 
     w*(0.648419685395586217984681 + 
        w*(0.452196867009616298571941 + 
           w*(0.371651133863068554240291 + 
              w*(0.32860784309392901699825 + 
                 w*(0.302354425634672539752731 + 
                    w*(0.285017852611712836182585 + 
                       w*(0.272940090886219734608815 + 
                        w*(0.264185174645675414204326 + 
                        w*(0.257629942234930169112297 + 
                        w*(0.25257672730254977798406 + 
                        w*(0.248568819726342542363224 + 
                        w*(0.245294721047639736371026 + 
                        w*(0.242535016745509439912893 + 
                        w*(0.240131243639121611881269 + 
                        w*(0.237966799943489229416831 + 
                        w*(0.235954779696462330935446 + 
                        w*(0.234029956041290960707096 + 
                        w*(0.232143339065715759679999 + 
                        w*(0.230258380102825285626612 + 
                        w*(0.228348256742232151582891 + 
                        w*(0.226393883462430267805038 + 
                        w*(0.224382419242756015826282 + 
                        w*(0.2223061215967177252663 + 
                        w*(0.220161445928088782138854 + 
                        w*(0.217948321166808082853286 + 
                        w*(0.215669553853982023082385 + 
                        w*(0.213330327149878379304031 + 
                        w*(0.210937771052290678843475 + 
                        w*(0.208500586950509676634869 + 
                        w*(0.206028714471339648851334 + 
                        w*(0.203533032028819495783199 + 
                        w*(0.201025084985861029012365 + 
                        w*(0.198516837154295178958399 + 
                        (0.196020442690960745941278 + 
                        0.19354803642357645356947*w)*w)))))))))))))
                        )))))))))))))))))))))
                    \end{verbatim}
and gives a directly usable implementation in C/C++ that can be pasted into a simulation code. Full details of how this was generated are given in Appendix B and are also available on-line. 

\subsection{Tail model for the stable quantile}
The power series computed thus far is a power series about the median, and as shown in the example precision plot for the Gaussian, loses precision in the tail. In each case we need to augment the quantile model by producing a separate model for the tail. We can of course add more and more terms to the power series, but ultimately we are confronted by the fact that usually such series should diverge as $|2u-1|\rightarrow 1$ and a polynomial truncation of the series is not enough.

There is an interesting mathematical question here. Should one try to develop a general theory based on some asymptotic analysis of the characteristic quantile equation? Or should one employ known information about particular cases? In this note we will take the latter approach, as there is a great deal of useful information available.

In the symmetric stable case the relevant asymptotic series for the distribution function were established over 50 years ago by Bergstr\"{o}m \cite{bergstrom}, based on earlier work on Laplace transforms by Pollard and others \cite{pollard}. An easily accessible paper \cite{matsui} has a formula (Eqn 2.9 of \cite{matsui}) whose integral gives, for $x\rightarrow \infty$, and $\alpha \neq 2$, and after some simplification
\begin{equation}
F(x) \sim 1- \frac{1}{\pi}\sum_{k=1}^N\frac{\Gamma(k\alpha)}{k!}(-1)^{k-1}\sin\left(\frac{k\pi\alpha}{2}\right)\frac{1}{x^{k\alpha}}
\end{equation}
We have found that the formal inversion of this series with $N=4$ provides a satisfactory tail model with a reasonably kinkless join to the power series. The result takes the form, for $\alpha < 2$,
\begin{equation}
w(u) \sim \left\{\frac{c_{-1}}{(1-u)} + c_0 + c_1(1-u)+c_2(1-u)^2\right\}^{(1/\alpha)}
\end{equation}
where the coefficients are given by
\begin{equation}
\begin{split}
c_{-1} &= \frac{\Gamma (\alpha ) \sin \left(\frac{\pi  \alpha }{2}\right)}{\pi} \\
c_0 &= -\frac{\cos \left(\frac{\pi  \alpha }{2}\right) \Gamma (2 \alpha
   )}{\Gamma (\alpha )}\\
c_1 &=\frac{\pi  \csc ^2\left(\frac{\pi  \alpha }{2}\right) \left(2 \Gamma
   (\alpha ) \Gamma (3 \alpha ) \sin \left(\frac{3 \pi  \alpha
   }{2}\right)-3 \csc \left(\frac{\pi  \alpha }{2}\right) \Gamma (2
   \alpha )^2 \sin ^2(\pi  \alpha )\right)}{12 \Gamma (\alpha )^3}\\
c_2 &= -\frac{\pi ^2 \cot \left(\frac{\pi  \alpha }{2}\right) \csc
   \left(\frac{\pi  \alpha }{2}\right) \left(6 (\cos (\pi  \alpha
   )+1) \Gamma (2 \alpha )^3-3 (2 \cos (\pi  \alpha )+1) \Gamma
   (\alpha ) \Gamma (3 \alpha ) \Gamma (2 \alpha )+\cos (\pi  \alpha
   ) \Gamma (\alpha )^2 \Gamma (4 \alpha )\right)}{6 \Gamma (\alpha
   )^5}
\end{split}
\end{equation}
The testing of such representations for intermediate $\alpha$ requires comparison with specialist models for the stable distribution. In this case we made a comparison of the results with those published on the web by J. Nolan, at
\begin{verbatim}
http://academic2.american.edu/~jpnolan/stable/quantile.dat
\end{verbatim}
We found that for $0.1 < u < 0.9$ the relative error between our results and Nolan's easily is less than $10^{-6}$, giving high confidence in the power series. With the simple tail model the precision over the entire range is as shown in the figure.

\subsection{Non-analytic density functions}
In the detailed analysis given thus far, we have considered the {\it analytic} case. By no means all densities of interest fall into this category, and we must also treat cases where not all of the characteristic moments exist. For example, consider the (symmetric) variance gamma (VG) case. Taking the limit $\delta \rightarrow 0_+$ in the generalized hyperbolic model gives us the characteristic function
\begin{equation}
\psi(t) = \left(\frac{\alpha^2}{\alpha^2+t^2} \right)^{\lambda}
\end{equation}
and the density
\begin{equation}
f(x) = \left(\frac{\alpha}{2} \right)^{\lambda+1/2}\frac{1}{\sqrt{\pi}\Gamma(\lambda)}|x|^{\lambda-1/2}K_{\lambda-1/2}(\alpha|x|)
\end{equation}
The full VG model falls into the category of exponential asymmetry, with characteristic function
\begin{equation}
\psi(t) = \left(\frac{\alpha^2-\beta^2}{\alpha^2+(t-i\beta)^2} \right)^{\lambda}
\end{equation}
A careful expansion of the density reveals that it comprises one power series in $x^2$ and a second power series in $x^2$ times $x^{(2\lambda-1)}$, so that a generalization of the methods developed here is needed for general $\lambda$. There are logarithmic contributions when $\lambda=1/2$. In general $\psi \sim t^{-2\lambda}$ as $t\rightarrow \infty$ so we can also see that not all the characteristic moments exist. One approach to VG is to exploit a non-uniform base distribution, as discussed in \cite{qmtwo} - in particular an exponential base is a convenient starting point for VG, hyperbolic and normal distributions.
\section{Conclusions}
We have shown in principle how to establish a power series about the median for quantile functions characterized by a characteristic function linked to a smooth but possibly non-explicit density. This has been elucidated in sufficient detail for symmetric distributions based on a detailed symbolic solution of a non-linear integro-differential equation. Further work is in progress to treat asymmetric cases and characteristic functions corresponding to non-smooth densities, as well as tail representations.

\subsection*{Acknowledgements}
We wish to acknowledge useful conversations or correspondence with G. Steinbrecher, J.P. Nolan and D. Scott. J. McCabe is supported by the UK EPSRC.

\section*{Appendix A: Mathematica code for the symbolic recursion}
The actual program used for the symmetric case was as follows. Here we work out every other term, but start the recursion carefully. First we give the starting values:
\begin{verbatim}
PP[2, x_, w_] := 
 1/(2 Pi) Integrate[
   I t \[CapitalPhi][t] Exp[-I t w], {t, -Infinity, Infinity}]

PP[3, x, w] = 
  Expand[3*x*PP[2, x, w]*PP[2, x, w] + 
    x^2 D[PP[2, x, w], x] PP[2, x, w] + D[PP[2, x, w], w]]  ;
   \end{verbatim}
Next we give a second order iteration based on a repeated application of the first order form:   
\begin{verbatim}
PP[n_, x_, w_] := 
 PP[n, x, w] = 
  Expand[n*x*
     PP[2, x, 
      w]*((n - 1)*x*PP[2, x, w]*PP[n - 2, x, w] + 
       x^2 D[PP[n - 2, x, w], x] PP[2, x, w] + 
       D[PP[n - 2, x, w], w]) + 
    x^2 D[((n - 1)*x*PP[2, x, w]*PP[n - 2, x, w] + 
        x^2 D[PP[n - 2, x, w], x] PP[2, x, w] + 
        D[PP[n - 2, x, w], w]), x] PP[2, x, w] + 
    D[((n - 1)*x*PP[2, x, w]*PP[n - 2, x, w] + 
       x^2 D[PP[n - 2, x, w], x] PP[2, x, w] + 
       D[PP[n - 2, x, w], w]), w]]
\end{verbatim}

Next we give the rules that set to zero terms that must vanish by symmetry:
\begin{verbatim}
rulesa = Table[
   Integrate[
     I*t^(4 k + 1)*\[CapitalPhi][t], {t, -Infinity, Infinity}] -> 
    0, {k, 0, 100}];
rulesb = Table[
   Integrate[-I*t^(4 k - 1)*\[CapitalPhi][t], {t, -Infinity, 
      Infinity}] -> 0, {k, 1, 100}];
rules = Join[rulesa, rulesb];\end{verbatim}
Then we give some results to write integrals in terms of the $E_k$.
\begin{verbatim}
srules = Join[
   Table[Integrate[
      t^(2 k) \[CapitalPhi][t], {t, -Infinity, Infinity}] -> 
     2 Pi EE[k], {k, 1, 200}],  
   Table[Integrate[-t^(2 k) \[CapitalPhi][t], {t, -Infinity, 
       Infinity}] -> -2 Pi EE[k], {k, 2, 200}]];\end{verbatim}
We initialize an array for solution:
\begin{verbatim}
parr = Array[p, 200];
\end{verbatim}
We run the iteration as far as time allows, writing a file every ten steps, or every iteration after 60, to secure progress made so far.
\begin{verbatim}
Do[dummya = PP[k, x, w] /. w -> 0; 
 dummyb = (dummya /. rules) /. srules; p[k] = Collect[dummyb, x]; 
 If[k >= 7, (PP[k - 2, x, w] = 0)];
 Print[k, " Memory in Use =  ", N[MemoryInUse[]/10^6]];
 If[Mod[k + 1, 10] == 0 || k > 60 , 
  Save[ToString[k] <> "pcoeff", parr]], {k, 3, 101, 2}] \end{verbatim}
In our case we ran the code on a 2.8GHz Mac Pro and secured terms up to $p_{71}$. We would be interested to hear about more efficient solutions. However, this many terms is sufficient for some serious practical applications, and we worked with the file {\tt 71pcoeff} for the computed examples. 

The {\it Mathematica} notebook {\tt CharacteristicQuantileSeriesGen} to generate the coefficients and the output files {\tt xxpcoeffs} are available from the links at the web site
\begin{verbatim}
www.mth.kcl.ac.uk/~shaww/web_page/papers/charquantiles/
\end{verbatim}
\section*{Appendix B: Generation of C/C++ code for the stable quantile}
Here we show how to generate C/C++ code in the late migration approach for the case of the stable quantile. The {\it Mathematica} notebook {\tt CharacteristicQuantileExamples} to do this is available from the web site
\begin{verbatim}
www.mth.kcl.ac.uk/~shaww/web_page/papers/charquantiles/
\end{verbatim}
The operations needed are as follows. First we load the general series data:
\begin{verbatim}
<< "71pcoeff"
\end{verbatim}
Next we define the characteristic moments, $w'(0)$ and some rules:
\begin{verbatim}
g[k_] = 2 Integrate[t^(2 k)*Exp[-Abs[t]^\[Alpha]], {t, 0, Infinity}, 
   Assumptions -> {\[Alpha] > 0, k > -1/2}];
stablerules = Table[EE[k] -> 1/(2 Pi) g[k], {k, 1, 60}];
wdash = 1/( 1/(2 Pi) g[0]);
\end{verbatim}
The definitions of the rules {\tt rulesa, rulesb, rules, srules} given in Appendix A are also re-loaded, and the series is then constructed in two steps:
\begin{verbatim}
series = Table[(((parr[[2 m + 1]] x/(2 m + 1)! ) /. rules) /. 
      srules /. stablerules), {m, 1, 35}];
StableSeries = series /. x -> wdash;
\end{verbatim}
\subsection*{Creating the Gaussian example}
This is now easily done:
\begin{verbatim}
GaussStableSeries = (series /. x -> wdash /. \[Alpha] -> 2);
wdashGauss = (wdash /. \[Alpha] -> 2);
GaussStableQuantile[u_] = 
  1/Sqrt[2] (wdashGauss*(u - 1/2) + 
     Sum[GaussStableSeries[[k]]*(wdashGauss*(u - 1/2))^(2 k + 1), {k, 
       1, 35}]);
\end{verbatim}

\subsection*{Creating the $\alpha=3/2$ example}
This is now easily done:
\begin{verbatim}
threehalfStableSeries = (series /. x -> wdash /. \[Alpha] -> 3/2);
wdashthreehalf = (wdash /. \[Alpha] -> 3/2);
ThreeHalfStableQuantile[
   u_] = (wdashthreehalf*(u - 1/2) + 
    Sum[threehalfStableSeries[[
       k]]*(wdashthreehalf*(u - 1/2))^(2 k + 1), {k, 1, 35}]);
\end{verbatim}
The function {\tt ThreeHalfStableQuantile} may then be used within {\it Mathematica}. Within the late migration approach, we convert to C/C++ by first creating a suitable form for export. The following code accomplishes three things
\begin{itemize}
\item works in terms of $2u-1$ in order to get radius of convergence unity and better control of the scale of the coefficients;
\item uses variables $v=2u-1$, $w=v^2$;
\item converts to nested multiplication form.
\end{itemize}
\begin{verbatim}
exportThreeHalf = 
  HornerForm[
    N[(wdashthreehalf*v/2 + 
       Sum[threehalfStableSeries[[
          k]]*(wdashthreehalf*(v/2))^(2 k + 1), {k, 1, 35}]), 24]] /. 
   v^2 -> w;\end{verbatim}
Then to generate the C/C++ code listing in Section 7.3 , we merely apply
\begin{verbatim}
CForm[exportThreeHalf]
\end{verbatim}
to obtain the results shown. 
\end{document}